\documentclass[aps, prb,superscriptaddress,amsmath,amssymb,preprintnumbers,twocolumn,floatfix,showpacs,showkeys,10pt]{revtex4-1}
 \usepackage{amssymb} \usepackage{epsfig} \usepackage{color} \usepackage{empheq}  \usepackage{tabularx}
 \makeatother
 \begin{document}
  \title{Magneto-photoluminescence in GaAs/AlAs core-multishell nanowires: a theoretical investigation}
   \author{Fabrizio~Buscemi}
\email{fabrizio.buscemi@unimore.it} \affiliation{Department of Physics, Informatics, and Mathematics, University of  Modena and  Reggio Emilia, Via Campi 213/A, I-41125 Modena, Italy}
   \author{Miquel~Royo }
\affiliation{Departament de Quim\'{i}ca F\'{i}sica i Anal\'{i}tica, Universitat Jaume I,E-12080 Castell\'{o}, Spain}
 \affiliation{CNR-NANO S3, Institute for Nanoscience, Via Campi 213/A, 41125}
  \author{Andrea~Bertoni }
 \affiliation{CNR-NANO S3, Institute for Nanoscience, Via Campi 213/A, 41125}
  \author{Guido~Goldoni  }
 \affiliation{Department of Physics, Informatics, and Mathematics, University of  Modena and  Reggio Emilia, Via Campi 213/A, I-41125 Modena, Italy}
\affiliation{CNR-NANO S3, Institute for Nanoscience, Via Campi 213/A, 41125}
\date{\today}

\newcommand{\guido}[1]{\textbf{GUIDO}: \textit{#1}}

 \begin{abstract} 
 The magneto-photoluminescence in modulation doped core-multishell nanowires is predicted as a function of photo-excitation intensity in non-perturbative transverse magnetic fields. We use a self-consistent field approach within the effective mass approximation to determine the photoexcited electron and hole populations, including the complex composition and anisotropic geometry of the nano-material. The evolution of the photoluminescence is analyzed as a function of \textit{i}) photo-excitation power, \textit{ii}) magnetic field intensity, \textit{iii}) type of doping, and \textit{iv}) anisotropy with respect to field orientation.
 \end{abstract}

 \maketitle

 \section{Introduction}

In the last decade, semiconductor nanowires (NWs)  have received increasing attention owing to their potential applications in nanotechnology.~\cite{Lieber2007}  
They represent promising candidates for  a wide range of novel  ultrafast electronic and optoelectronic nanodevices including high electron mobility transistors,~\cite{Cui2003}  photovoltaic cells,~\cite{Czaban2009}  single-photon emitters,~\cite{Holmes2014} and lasers.~\cite{Qian} 
Critical steps have been taken in these directions in the last years. 
Long single-crystal, defect-free cores,~\cite{Shtrikamn2009,Shtrikamn2009b} selective radial doping,~\cite{Tomioka2010} lateral overgrowth with high-quality interfaces~\cite{Tambe2008} have been successfully realized. Most importantly, complex radial modulation doped heterostructures are now being engineered in core-multishell NWs.~\cite{Funk2013, Fick2013, Jadc2014,Shi2015} 
More complex than their planar counterparts, radial heterostructures in these systems may host high-mobility electron/hole gases with an inhomogeneous localization in the section of the NW, which strongly depends on doping density and gate potentials.~\cite{Wong2011, Bertoni2011,Tomioka2012}

Magnetic states of such axial electron gases have been investigated in cylindrical systems~\cite{Hiroshi1993,Ferrari2008,Bellucci2010} or in wrapped heterojunctions with prismatic symmetry.~\cite{Ferrari2009, Royo2013,Royo2015} In the Quantum Hall regime, the magnetic field competes with the self-consistent confinement field.
Recent self-consistent field calculations of realistic GaAs/AlGaAs core-shell NWs in transverse fields~\cite{Royo2013} showed that at high carrier density the carrier-carrier repulsion results in complex magnetic bands, corresponding to inhomogeneous charge localization at the prismatic hetero-interfaces of the nanomaterial. 

Intraband optical spectroscopy is a key component in the investigation of properties of single semiconductor NWs and NW  heterostructures.~\cite{Titova2007,Hoang2007,Boland2015,Lei2010, Fick2013, Jadc2014, Shi2015} 
In particular, photoluminescence (PL) has been successfully used in bare and core-multishell (CMS) NWs in order to analyze the dynamics of photo-excited electron-hole plasmas~\cite{Titova2007,Hoang2007, Boland2015} and the quantum confinement effects.~\cite{Fick2013,Jadc2014,Shi2015}
Magneto-PL in  GaAs/AlAs CMS NWs has been applied to undoped~\cite{Plochocka2013} and doped samples,\cite{Jadc2014} 
where the magnetic field dependence of peaks ascribed to confined carrier emission has been examined.

In this work, we consider magneto-PL in modulation doped GaAs/AlAs CMS NWs with an inversion or an accumulation layer at the inner GaAs/AlAs hetero-interface in a transverse magnetic field.  
Using a self-consistent envelope-function approach, we predict the magneto-PL at different excitation powers and identify non-linear effects due to filling and restructuring of the free-carrier gas. 
Furthermore, we investigate the effects of the type of doping and of the structure anisotropy on the magneto-PL. 

The article is organized as follows.
In Sec.~\ref{approach}, we describe the physical and numerical modeling of the modulation doped CMS NW used to evaluate the hole and electron states and from these the PL spectra.
The results obtained from our numerical calculations are discussed in Sec.~\ref{res}.
First, the prototype nanostructure under investigation and its relevant physical parameters are presented in subsection~\ref{structuredetails}.
Then, subsection~\ref{respdop} illustrates predictions for the $p$- and $n$-doped system at various regimes of photo-excitation power.
Anisotropy with respect to field orientation is examined in subsection~\ref{subani}.
We summarize our results in Sec.~\ref{conclu}.

\section{Theoretical approach} \label{approach}

Electron and hole states are determined by a self-consistent Schr\"odinger-Poisson approach.~\cite{Bertoni2011,Wong2011, Royo2013} 
Within a single parabolic band effective-mass approximation, the Hamiltonians for holes ($h$, upper sign) and electrons ($e$, lower sign) in an external magnetic field
\footnote{The small Zeeman energy ($<0.3$~meV at the largest field used) is below the thermal linewidth\cite{Jadc2014} and has been neglected.}
are given by
\begin{widetext}
\begin{equation} \label{sch1}
\left[ \left( \textbf{P} \mp q\textbf{A}(\textbf{R}) \right)
       \frac{1}{2m_{h(e)}^{\ast}(\textbf{R})}
       \left( \textbf{P} \mp q\textbf{A}(\textbf{R}) \right)
\mp E_{V(C)}(\textbf{R}) \pm qV(\textbf{R}) \right]
\end{equation}
\end{widetext}
respectively, where \textbf{R}=$(x,y,z)$ represents a 3D coordinate, \textbf{P} is the conjugate momentum operator, $q$ is the positive unit charge,
\textbf{A}(\textbf{R}) is the magnetic vector potential, $m^{\ast}_{e(h)}(\textbf{R})$ is the material-dependent effective mass of electrons (holes),
and $E_V(\textbf{R})$, $E_C(\textbf{R})$, and $V(\textbf{R})$ are the local valence-band edge, the local conduction-band edge, and the electrostatic potential generated by the free carriers and the dopants, respectively.  
Electron-hole interactions are taken into account within a mean-field approximation through the self-consistent potential $V(\textbf{R})$, i.e., 'excitonic' effects are not included in our calculations, as they are expected to be negligible at densities larger than $\approx10^6$~electron/cm\cite{Rossi1996, Sedlmaier}
\footnote{We do not include the exchange and correlations effects in the mean field since they turn out to be negligible.~\cite{Jogai2002,Bertoni2011,Wong2011}} 

\begin{figure}[htpb]
  \begin{center} 
    \includegraphics*[width=0.7\linewidth]{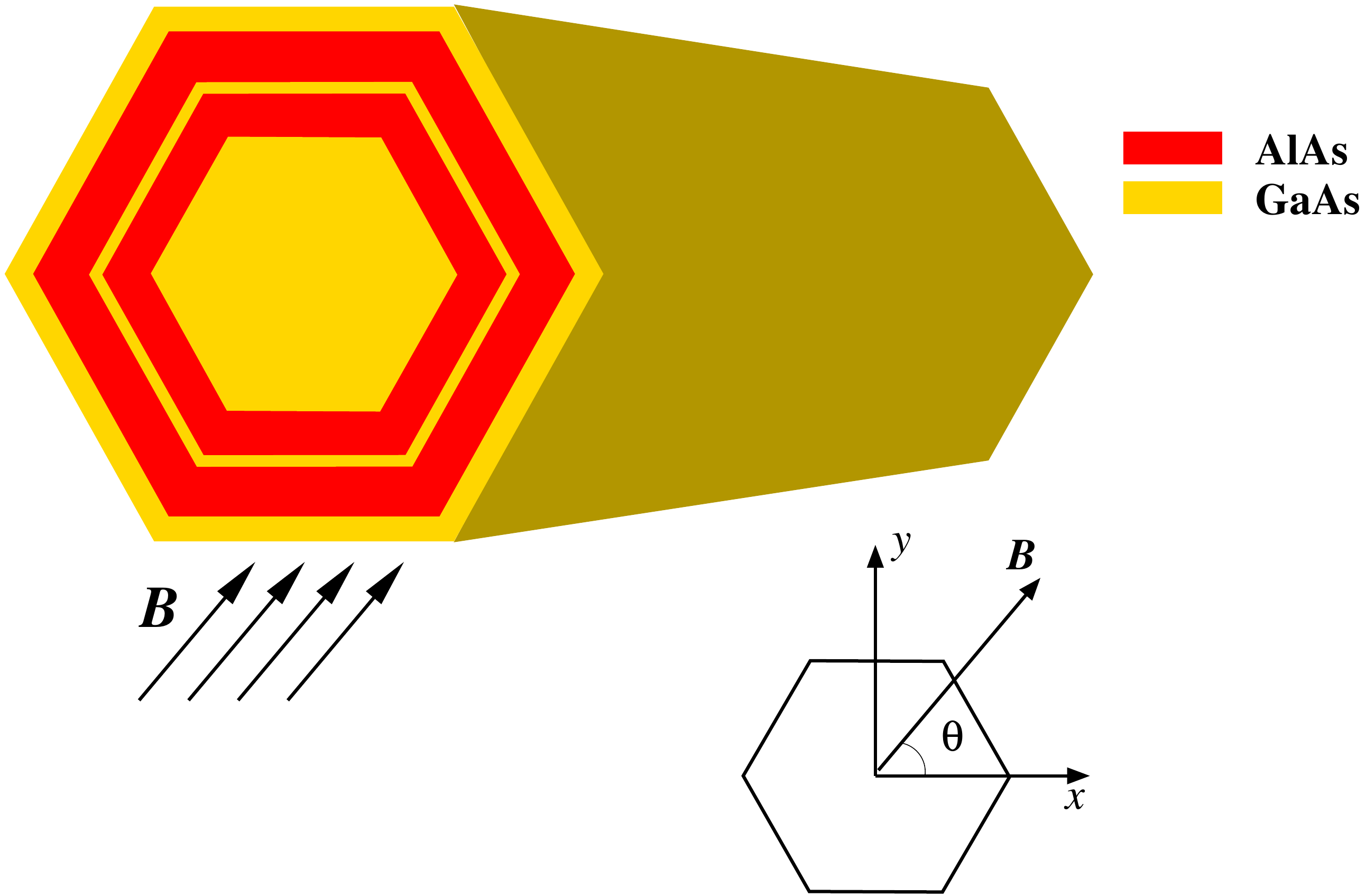}
    \caption{ \label{struct} Sketch of the GaAs/AlAs core-multi shell NW investigated in this work. The GaAs core is overgrown with AlAs/GaAs/AlAs layers forming a narrow quantum well and a GaAs capping layer. The GaAs quantum well serves as a doping layer with a homogeneously distributed density of donors or acceptors $n_D $ or $n_A)$, respectively. The transverse magnetic field forms an angle $\theta$ with the $x$-axis.}
  \end{center}
\end{figure}

The transverse magnetic field is described in the Landau gauge $\textbf{A}(\textbf{R})=B(y\cos \theta -x \sin \theta)\hat{z}$,
where $\theta$ denotes the angle between the  field and the $x$-axis (see Fig.~\ref{struct}). We shall consider two possible orientations of the field: \emph{i)} perpendicular to the facets ($\theta =\pi/2$) and \emph{ii)} along the maximal diameter
($\theta=\pi/3$).

Since the field does not break the translational invariance, the 3D wavefunctions can be factorized as 
$\Psi^{e(h)}_n(\textbf{R})=e^{ikz}\psi^{e(h)}_{n,k}(\textbf{r})$, where $\textbf{r}=(x,y)$ and $k$ the wavevector along the  nanowire axis. Then the 2D wavefunctions $\psi^{e(h)}_{n,k}(\textbf{r})$ are solutions of the Schr\"odinger equations
\begin{widetext}
\begin{eqnarray} \label{sch2}
& &\left[ -\frac{\hbar^2}{2} \nabla_{\textbf{r}} \frac{1}{m_{h,\perp}^{\ast}(\textbf{r})} \nabla_{\textbf{r}} +\frac{1}{2}m_{h,\parallel}^{\ast}(\textbf{r}) \omega_{c,h}^2(\textbf{r}) \left(y\cos \theta -x \sin \theta-kl^2_B \right)^2 -E_V(\textbf{r})+qV(\textbf{r}) \right]\psi^h_{n,k}(\textbf{r})=-E^h_{n,k}\psi^h_{n,k}(\textbf{r}) \nonumber \\
& &\left[ -\frac{\hbar^2}{2} \nabla_{\textbf{r}} \frac{1}{m_e^{\ast}(\textbf{r})} \nabla_{\textbf{r}} +\frac{1}{2}m_e^{\ast}(\textbf{r}) \omega_{c,e}^2(\textbf{r}) \left(y\cos \theta -x \sin \theta+kl^2_B \right)^2+E_C(\textbf{r})-qV(\textbf{r})\ \right]\psi^e_{n,k}(\textbf{r})=E^e_{n,k}\psi^e_{n,k}(\textbf{r}) ,
\end{eqnarray}
\end{widetext}
where $\omega_{c,e(h)}(\textbf{r})=qB/m_{e(h,\parallel)}^{\ast}(\textbf{r})$ is the hole (electron) cyclotron frequency, $l_B=\sqrt{\hbar/(qB)}$ is the magnetic length, and $E^{h(e)}_{n,k}$ is the energy of the hole (electron) state. Note that distinguish between the hole effective mass in the section of the CSNW, $m_{h,\perp}^{\ast}$, and the hole effective mass along the wire, $m_{h,\parallel}^{\ast}$, to take into consideration the strong mass anisotropy of GaAs. The electron effective mass, $m_e^{\ast}$, is assumed to be isotropic.

Similarly to the familiar case of a planar Hall bar, in the above equations the effect of the magnetic field is represented by an harmonic potential displaced by $kl^2_B$ from the nanowire axis. 
Therefore, in general, $E^{h(e)}_{n,k}$ is not parabolic in $k$ at large magnetic fields. Note that Eqs.~\ref{sch2}) are coupled by the electrostatic field $V(\textbf{r})$ which is determined by both electron and hole densities, as shown below.

Equations ~\eqref{sch2} are numerically solved by the box integration method on a triangular grid with hexagonal elements.~\cite{Bertoni2011, Royo2013}  
At each magnetic fields the equations are solved on a grid of wavevectors in $[0,k_{\textrm{max}}]$.~\footnote{Due to  the  symmetry of the Hamiltonian in Eqs.~\eqref{sch2}, states with negative $k$ can be obtained from $(\textbf{r},k) \rightarrow (-\textbf{r},-k)$.}
The $k$-domain is discretized in steps $\Delta k=0.075$~nm$^{-1}$, while $k_{\textrm{max}}$ is taken sufficiently larger than the Fermi wave vector that its occupation is negligible. 
The hole and electron densities are evaluated at temperature $T$ from
\begin{eqnarray} \label{densit}
 & &n_h^{\circ}(\textbf{r},E_F)=2\sum_n \int_{-k_{\textrm{max}}}^{k_{\textrm{max}}} \frac{dk}{2\pi}  |\psi^h_{n,k}(\textbf{r})|^2 f\left(\frac{-E^h_{n,k}+E_F}{k_B\textrm{T}}\right) \nonumber \\
 & &n_e^{\circ}(\textbf{r},E_F)=2\sum_n \int_{-k_{\textrm{max}}}^{k_{\textrm{max}}} \frac{dk}{2\pi}  |\psi^e_{n,k}(\textbf{r})|^2 f\left(\frac{E^e_{n,k}-E_F}{k_B\textrm{T}}\right)
\end{eqnarray}
where the factor 2 accounts for spin degeneracy, and $f(x)=1/(1+e^{x})$
is the Fermi function. $E_F$ indicates the Fermi level which is determined by surface states and, therefore, by the type of doping. The charge densities in Eqs.~\eqref{densit} depend on the fully-ionized dopants, i.e., photo-excited carriers are not included here.
As shown elsewhere,~\cite{Bertoni2011,Royo2013} the symmetry and localization of $n_h^{\circ}$ and $n_e^{\circ}$ are strongly influenced by the level of doping, due to competing energy scales.~\cite{Royo2013, Jadc2014} 
The linear free charge density per unit length provided by the dopants  $n_{\textrm{l},e(h)}^{\circ}$ can finally be calculated from 
\begin{equation}
n_{\textrm{l},e(h)}^{\circ}=\int_A n_{e(h)}^{\circ}(\textbf{r},E_F) d\textbf{r}.
\end{equation}
Its dependence from $E_F$ is omitted for brevity.

The laser field also excites electron-hole pairs with a linear density $n^{\textrm{ph}}_{\textrm{l}}$. 
If the photo-excited density is not negligible with respect to the free charge induced by dopants, the self-consistent field should take these charges into account.
The  total electron (hole) linear density $n_{\textrm{l},e(h)}$ is
\begin{equation}
n_{\textrm{l},e(h)}= n_{\textrm{l},e(h)}^{\circ} +n^{\textrm{ph}}_{\textrm{l}}.
\end{equation}
In order take into account the photoexcited charge, we shift the Fermi level $E_F$ to $E_F^{e(h)}$  (in general $E_F^h\neq E_F^e$) until 
\begin{equation} \label{constr} 
\int_A n_{e(h)}(\textbf{r},E_F^{e(h)}) d\textbf{r}-n_{\textrm{l},e(h)}=0.
\end{equation}
This amounts to the charge neutrality condition in an undoped sample. In practice, we obtain $E_F^{e(h)}$ from \eqref{densit}-(\ref{constr}) by means of the bisection method.


Once the total electron and hole densities are computed, the electrostatic potential $V(\textbf{r})$ is obtained by solving the Poisson equation
\begin{equation} \label{Poisson}
\nabla_{\textbf{r}} \left[ \epsilon_r(\textbf{r}) \nabla_{\textbf{r}} V(\textbf{r})\right]=-\frac{q}{\epsilon_0} [n_h(\textbf{r})-n_e(\textbf{r}) +n_D(\textbf{r})-n_A(\textbf{r})],
\end{equation}
where $\epsilon_r(\textbf{r})$ and $\epsilon_0$ describe the position-dependent dielectric constant and the vacuum permittivity, while $ n_{D(A)} (\textbf{r})$ denotes the ionized donor (acceptor) density. 
Dirichlet boundary conditions are used to solve Eq.~\eqref{Poisson} with the potential on the domain boundaries set to zero.

%
The electrostatic potential is inserted into Eqs.~\eqref{sch2} and the whole procedure is iterated until self-consistency is achieved, that is, the relative variation of both the electron and hole density is $< 0.01$ at each point of the simulation domain. 

The PL spectra is computed under the assumption that carriers are thermalized in the lowest electronic levels by fast non-radiative processes. Thus, the intensity
of PL signal is given by~\cite{Bastard}
\begin{widetext}
\begin{equation} \label{PLI}
I_{\textrm{PL}}(\omega) \propto \sum_{n,m} \int_{-k_{\textrm{max}}}^{k_{\textrm{max}}} \frac{dk}{2\pi} \left|\langle\psi^e_{n,k}| \psi^h_{m,k}\rangle\right|^2   \times f\left(\frac{-E^h_{m,k}+E^h_F}{k\textrm{T}}\right) f\left(\frac{E^e_{n,k}-E^e_F}{k\textrm{T}}\right) \nonumber 
\Im \left[  \frac{1}{E^e_{n,k}-E^h_{m,k}-\hbar \omega -i \gamma}   \right],
\end{equation}
\end{widetext}
where $\langle\psi^e_{n,k}| \psi^h_{m,k}\rangle=\int d\textbf{r} {\psi^e_{n,k}}^{\ast}(\textbf{r}) \psi^h_{m,k}(\textbf{r})$, the product of the two Fermi distributions describes the occupation of the states
involved in the photoemission, and the last term accounts for a Lorentzian broadening of the peaks due to incoherent processes by means of a phenomenological parameter $\gamma$. As usual for interband transitions at optical frequencies, only \emph{vertical} transitions are taken into account.

\section{Results} \label{res}

\subsection{Structure details} \label{structuredetails}
\newcolumntype{C}[1]{>{\centering\arraybackslash}p{#1}}
We simulate a prototype NW with the same compositional parameters as in Ref.\onlinecite{Jadc2014}. A GaAs core of diameter 60 nm (facet to facet) is surrounded
by a 3 nm/1 nm/3 nm AlAs/GaAs/AlAs multilayer and a 13 nm GaAs capping layer. Donors or acceptors are uniformly distributed in the narrow GaAs quantum well. 
For $p$-doping we assume a Fermi level pining at 400~meV above the valence band edge,~\cite{Pash93}  while for $n$-doping we assume mid gap pinning. 
The PL phenomenological linewidth $\gamma = 0.5\,\mbox{meV}$ and all calculations have been performed at temperature T~$ =1.8$~K. 
\begin{table}
\centering
\begin{tabular}{p{2.5cm} C{2.5cm} C{2.5cm}}
  \hline \hline
     & GaAs &  AlAs \\  \hline  
$E_C-E_V$(meV)   & 1519  & 3020 \\  
   $m_e^{\ast}$ & 0.062  & 0.19\\  
   $m_{h,\parallel}^{\ast}$ & ~0.082  & 0.109\\  
   $m_{h,\perp}^{\ast}$ & 0.680 &  0.818  \\  \hline  \hline
\end{tabular}
\caption{ Material parameters used in the numerical calculations.~\cite{1997handbook}
The relative effective masses in the axial ($z$) and transverse ($\textbf{r}$) directions are given in units of the free electron mass.\label{tab1} }
\end{table}

Material parameters are reported in Tab.~\ref{tab1}. Hole masses are chosen as those of a planar quantum well grown along the [110] direction, which corresponds to the radial quantum well of the CSNW structure. Appropriate expressions of the effective masses along the quantization axis [110] of the quantum well and the in-plane direction (here corresponding to the in-wire direction [111]) are obtained, e.g., in Ref.~\onlinecite{Fishman} from the diagonalization of the Luttinger Hamiltonian. Note that the 'heavy' mass along the [110] direction is taken as isotropic in the section of the CSNW, why a much lighter mass describes the heavy hole dispersion along the wire. 

Note that in our procedure we do not include light hole states. We have checked that 
light-hole subbands (evaluated within a parabolic approximation with the light-hole mass in the [110] growth direction~\cite{Fishman}) lie well below than the Fermi level.
Clearly, light-hole states might still contribute to heavy-hole states via spin-orbit coupling which is neglected in the parabolic approximation assumed in this work. Unfortunately, spin-orbit coupled states have not been investigated for realistic models of CSNWs. However, from previous calculations in semiconductor quantum wires, we may expect these effects to be small for these weakly confined states.~\cite{Goldoni1997}

Note that the narrow GaAs QW acts here as a doping layer. 
Due to its narrow thickness, no quantum state is hosted by the QW itself, and all free charge accumulates in an inversion or accumulation layer, depending on the type of doping, in the core or at the inner hetero-interface, as we shall discuss below.

\subsection{Fan diagrams} \label{respdop}

We examine the evolution of the magneto-PL with a transverse magnetic field for $p$- and $n$-doped samples. 
We shall consider three levels of the photo-excited charge, namely, negligible, lower or  comparable to the free-charge density induced by a typical doping level.

\subsubsection{$p$-doping}

We first consider a $p$-doped sample with $n_A=2.5 \times 10^{19} \mbox{cm}^{-3}$, which corresponds to a linear free-hole density $n_{\textrm{l},h}^{\circ} = 8.5 \times 10^{6} \mbox{cm}^{-1}$. 
We first consider a low excitation density  $n^{\textrm{ph}}_{\textrm{l}}=10^{5}~\mbox{cm}^{-1} \ll n_{\textrm{l},h}^{\circ}$. 

\begin{widetext}
\phantom{aa}
\begin{figure}[htpb]
  \begin{center} 
   \includegraphics[width=0.45\linewidth]{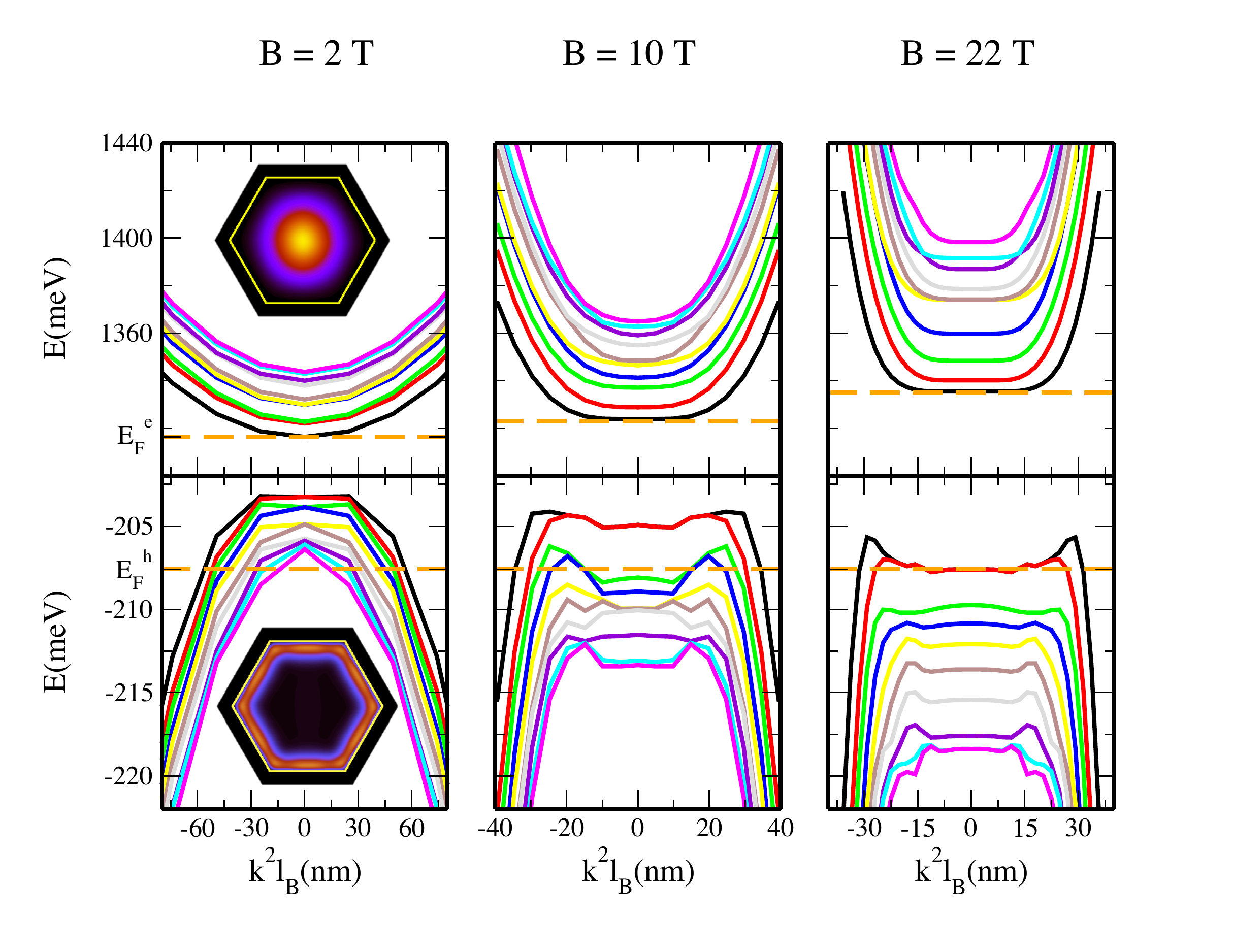}
  \includegraphics[width=0.45 \linewidth]{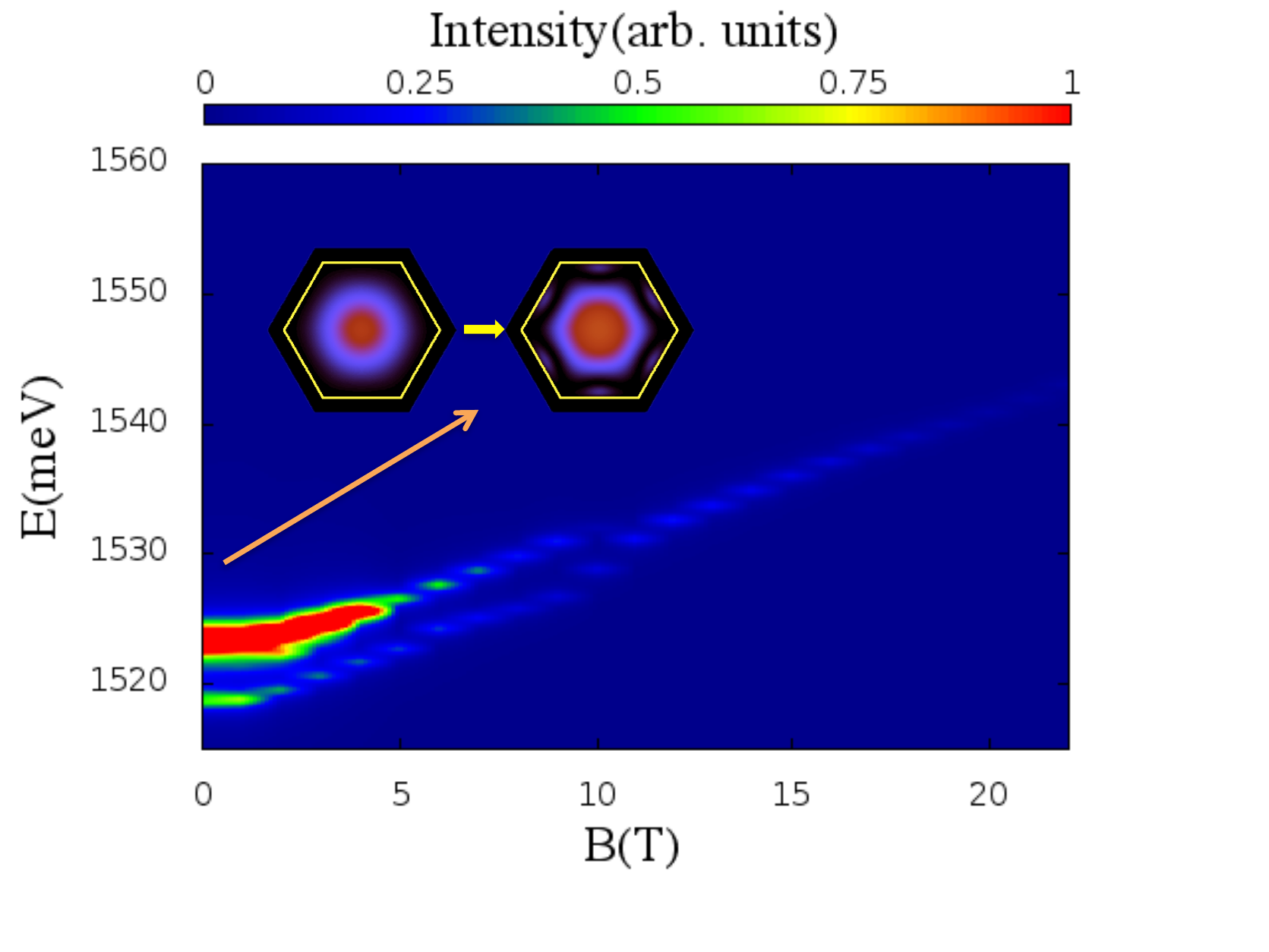}
         \caption{ \label{subba} Magnetic states and magneto-PL calculated for  $n_A=2.5 \times 10^{19}$~cm$^{-3}$ and $n^{\textrm{Ph}}_{\textrm{l}}=10^{5}$~cm$^{-1}$. Here $B$ is normal to the facets ($\theta =\pi/2$). Left:  electron (top) and hole (bottom) subbands \textrm{vs} $k l^2_B$  at selected fields $B$ as indicated.
The horizontal lines indicate the energy of the electron/hole Fermi level.
The hexagonal insets show the electron (top) and hole (bottom) charge density at $B=2$~T. Right: calculated PL spectra. The inset displays the squared envelope functions for the electron (left) and hole (right) state whose recombination yields the strong peak at  $E=1523$~meV when $B=0$. 
The regular intensity oscillations are a graphic artifact due to the course magnetic field grid.}
  \end{center}
\end{figure}
\end{widetext}

In Fig.~\ref{subba}, left panel, we show the dispersion of the magnetic levels for holes and electrons evaluated at selected $B$ values. The evolution of the magnetic subbands for conduction electrons resembles that of Landau levels and edge states in planar Hall bars.~\cite{Cardona} 
Indeed, as shown in the inset, the electron charge is localized in the core, since $n^{ph}_l$ is very small and it is not able to bend the conduction edge. Therefore, electron states are similar to those of a QW in a vertical field, here represented by the NW core, and bend up at increasing $k$, which correspond to edge states confined by the field on the lateral boundaries of the GaAs core.\cite{Royo2013} 

This is at difference with the dispersion of holes, which becomes flat at lower fields and develops non monotonous behavior at large $k$. 
Such a behavior is not due to the strong anisotropy of the hole mass but to the emergence of peculiar localized states.~\cite{Ferrari2008}
Indeed, holes, which here are mainly generated by doping so that $n_{l,h} \gg n_{l,e}$, create an accumulation layer near the hetero-interface, with an approximately tubular geometry, as shown in the inset. 
As discussed in Ref.~\onlinecite{Ferrari2008}, at sufficiently large fields the energy dispersion curves develop minima which correspond to states localized at the flanks of the tube, where the vertical component of the field with respect to the charge layer (and the associated confinement energy) is vanishing. 
Note from Fig.~2(left) that several hole levels are occupied, while electrons only occupy the lowest magnetic subband. 
The number of occupied subbands in the hole states decrease with increasing field. At intermediate fields ($B=10$~T), only 4 hole subbands are occupied and at 22 T only two hole magnetic subbands lie below or close to the hole Fermi level.

\begin{figure}[htpb]
  \begin{center} 
   \includegraphics*[width=\linewidth]{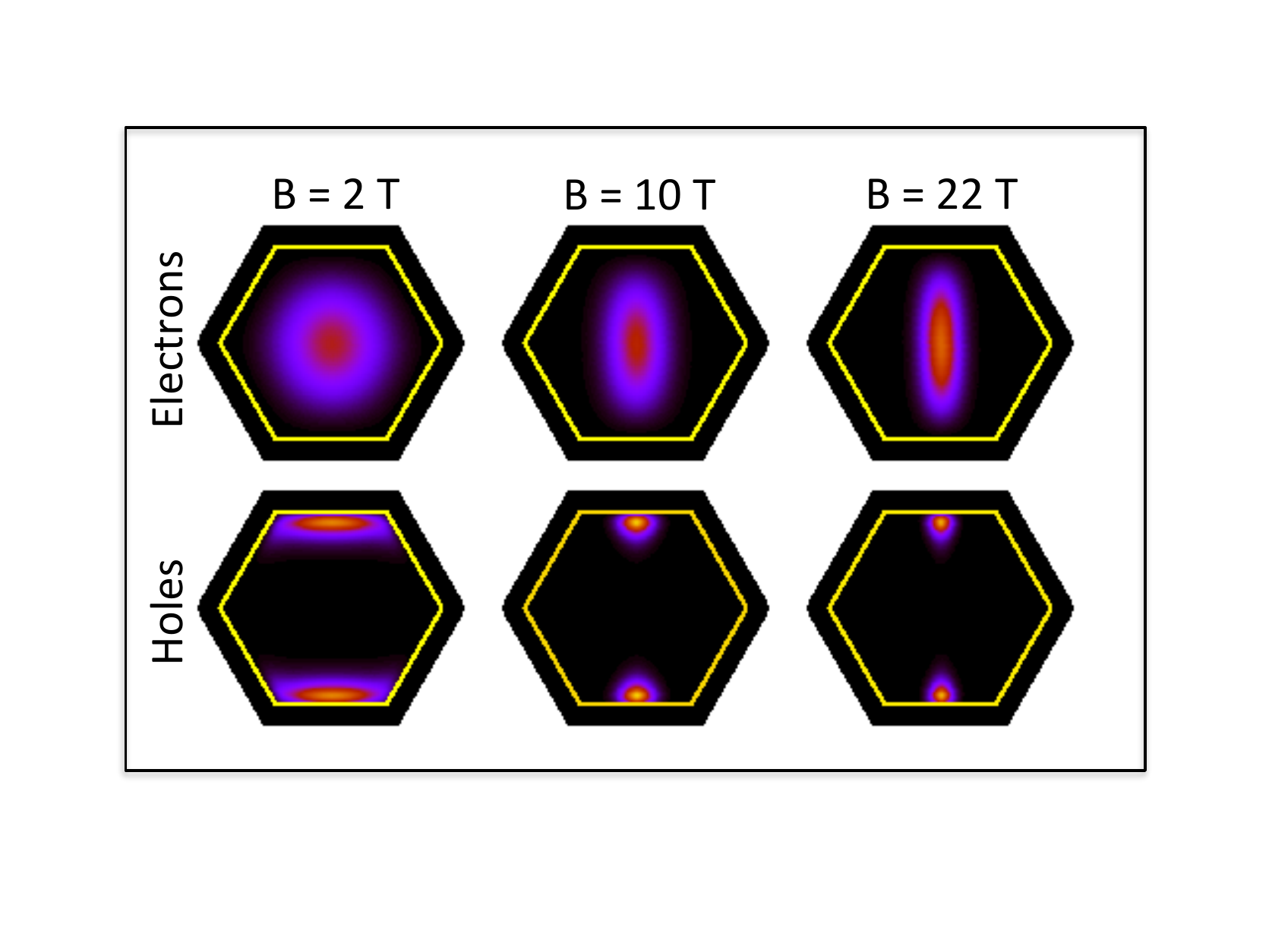}
            \caption{ \label{1e7states}             Square modulus of the lowest  $k=0$ electron (top) and hole (bottom) envelope function 
            at selected fields, as indicated, for the sample of Fig.~\ref{subba}. }
  \end{center}
\end{figure}

The evolution of the ground state $n=1$ and $k=0$ for electrons and holes is shown in Fig.~\ref{1e7states} at selected magnetic field intensities. While the electron state is delocalized over the core, the hole state is localized at the top and bottom of the structure (here the field is in the vertical direction). Clearly, the states extend laterally by about $l_B$.

The evolution of the calculated PL spectra with field is shown in the fan diagrams of Fig.~\ref{subba}, right panel. At zero field, the  spectrum exhibits two peaks. The lowest in energy, close to the band-gap of bulk GaAs at 1519~meV, is due to the electron and hole (e-h) ground state, coupling the states shown in Fig.~\ref{1e7states}. Due to the small overlap, this peak is weak. The second, much stronger peak originates from the recombination of  the electron ground state and the hole lowest excited state spreading over the center of  the GaAs core, hence the large oscillator strength (see the inset in the right panel of Fig.~\ref{subba}). 

\begin{figure}[htpb]
  \begin{center} 
   \includegraphics*[width=\linewidth]{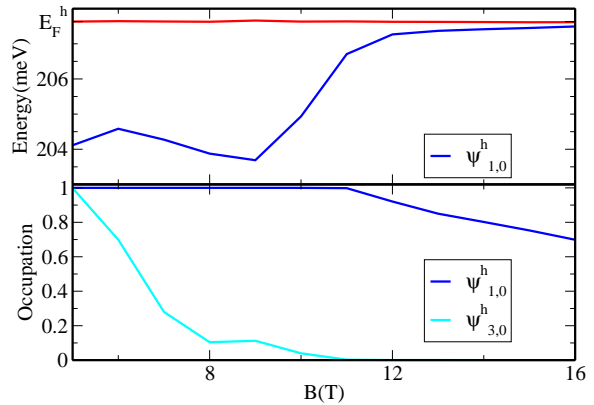}
            \caption{ \label{kink} Energy and subband occupation for the sample of Fig.~2. Top: Energy of the ground hole state at $k=0$ \textit{vs} $B$. The horizontal line indicates the hole Fermi level. Bottom: Occupation of $n=1$ (blue line) and $n=3$ (cyan line) hole states at $k=0$ as a function of the magnetic field.}
  \end{center}
\end{figure}

In the PL spectra in Fig.~\ref{subba} we can distinguish two different field regimes. At low fields, the peaks undergo a quadratic diamagnetic shift and progressively reduce their intensity. The stronger peak disappears at $B\sim$ 10 T, due to the decreasing occupation of the hole excited state ($n$=3). Indeed, even though the global hole density is almost independent from $B$, as the field increases the subbands flatten and accommodate a larger number of carriers in the lowest subband with $k\neq0$ as shown in Fig.~\ref{subba}(left). 
This leads to a  rearrangement  of the hole gas and the lowest energy subbands, and, in turn, the appearance of a kink in the emission line of the ground e-h state in the region around $B\sim$ 10~T. 
As shown in Fig.~\ref{kink}, just at the complete depletion of the $n=3$ hole level the gas redistribution affects the energy of the ground hole state at $k=0$  which increases by $\sim$4meV. Such an effect is related to the Fermi level pinning, that is $E_F^h$ goes from being pinned at the $n=3$ subband to the flat $n=1,2$ ones 
as found in other works.~\cite{Royo2013}
Note that in the present calculation PL spectra originate from the recombination of the hole and electron states with $k = 0$, since this is the only occupied state in the conduction band. 
At larger fields, when the dispersion flattens, also finite wavevectors are occupied. However, such states are localized on opposite sides of the structures for electrons and holes with the same $k$. Hence, the  overlap decreases and, as a consequence, the PL intensity decreases as well.

In the high field regime, the emission energy of the e-h ground state increases linearly with the field, as expected.~\cite{Jadc2014}
The peak also reduces its intensity as a consequence of both the decrease of the overlap integral between hole and electron states (see Fig.~\ref{1e7states}) and the decreasing occupation of the electron ground state at $k=0$.

At $n^{\textrm{ph}}_{\textrm{l}}=10^{6}$~cm$^{-1}$ the photo-excited hole density is of the same order of the free-charge density ($8.3\times10^{6}$~cm$^{-1}$ at $B=0$ T, $8.4\times10^{6}$~cm$^{-1}$ at $B=22$~T). Figure~\ref{subbaPh8}  displays the subband dispersion and the PL spectra for this configuration. 
The overall dispersion is similar to the previous case even though electrons now occupy a few excited subbands at the lowest field.  
The PL spectrum at $B=0$ exhibits four peaks, with the one due to e-h ground state  at  about 1520~meV being very weak.  Additional stronger peaks at 1523 meV, 1526 and 1530~meV are due to transitions between excited states. 
All emission lines exhibit a diamagnetic shift at low fields and a linear one at high magnetic fields. 
At increasing fields, the number of emission lines decreases since the occupation of  the upper electron levels vanishes. At $B=10$ only two peaks are visible, one related to  the e-h ground state and the other one originating from the recombination of an electron in the ground state and a hole in the excited state ($n=3$) spreading over the GaAs core (see the inset of the right panel of  Fig.~\ref{subbaPh8}). As $B$ increases, the $n=3$ subband gets closer to the lowest one in the $k$-interval around $k=0$. 
This is due to the fact that the $n=3$ hole state  develops lobes localized on opposite sides of the structure where  potential energy is lower with respect to the GaAs core.
Therefore the  two peaks approach each other and merge in an unique peak whose emission energy increases linearly with $B$.

\begin{widetext}
\phantom{aa}
 \begin{figure}[htpb]
  \begin{center} 
   \includegraphics[width=0.45\linewidth]{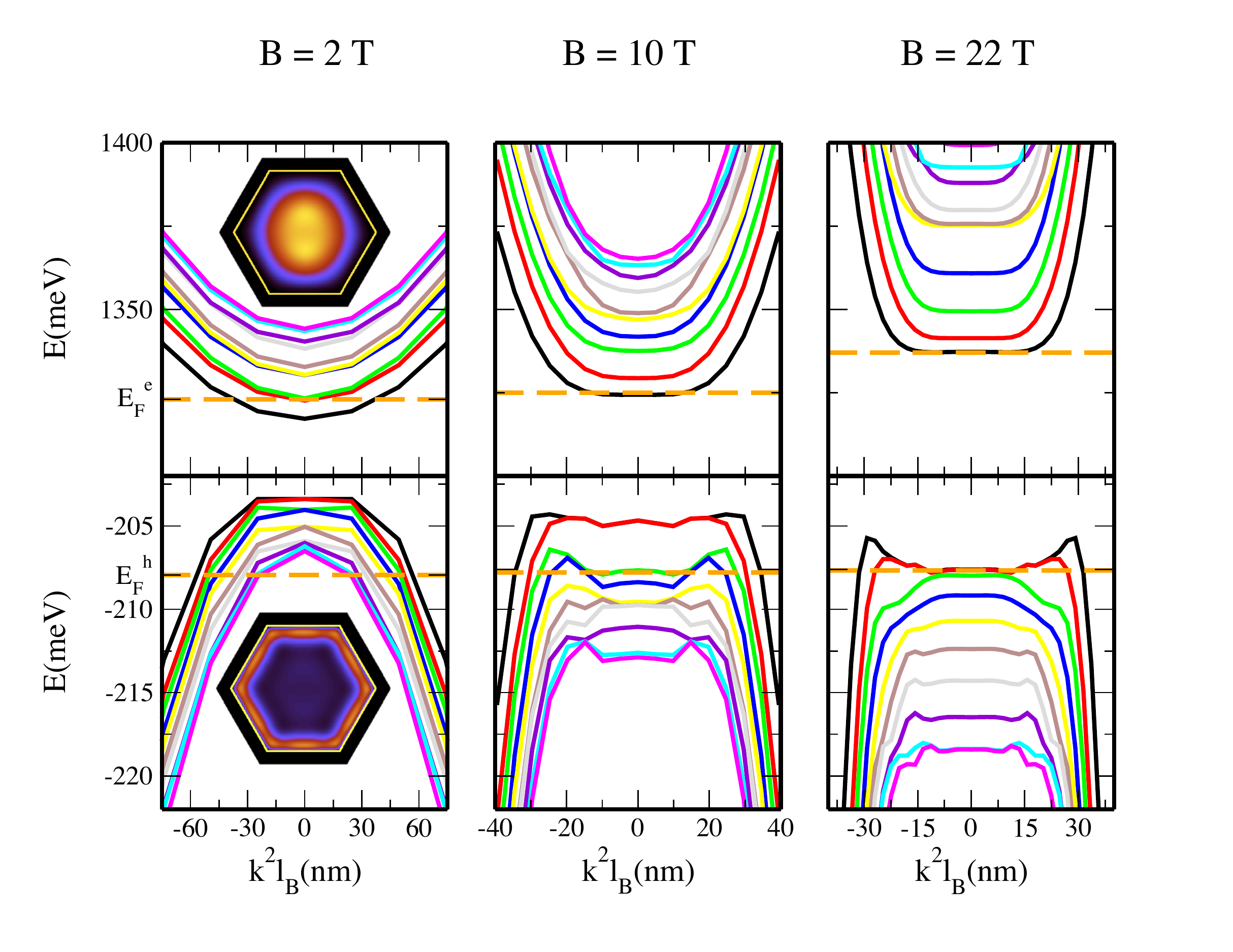}
   \includegraphics*[width=0.45\linewidth]{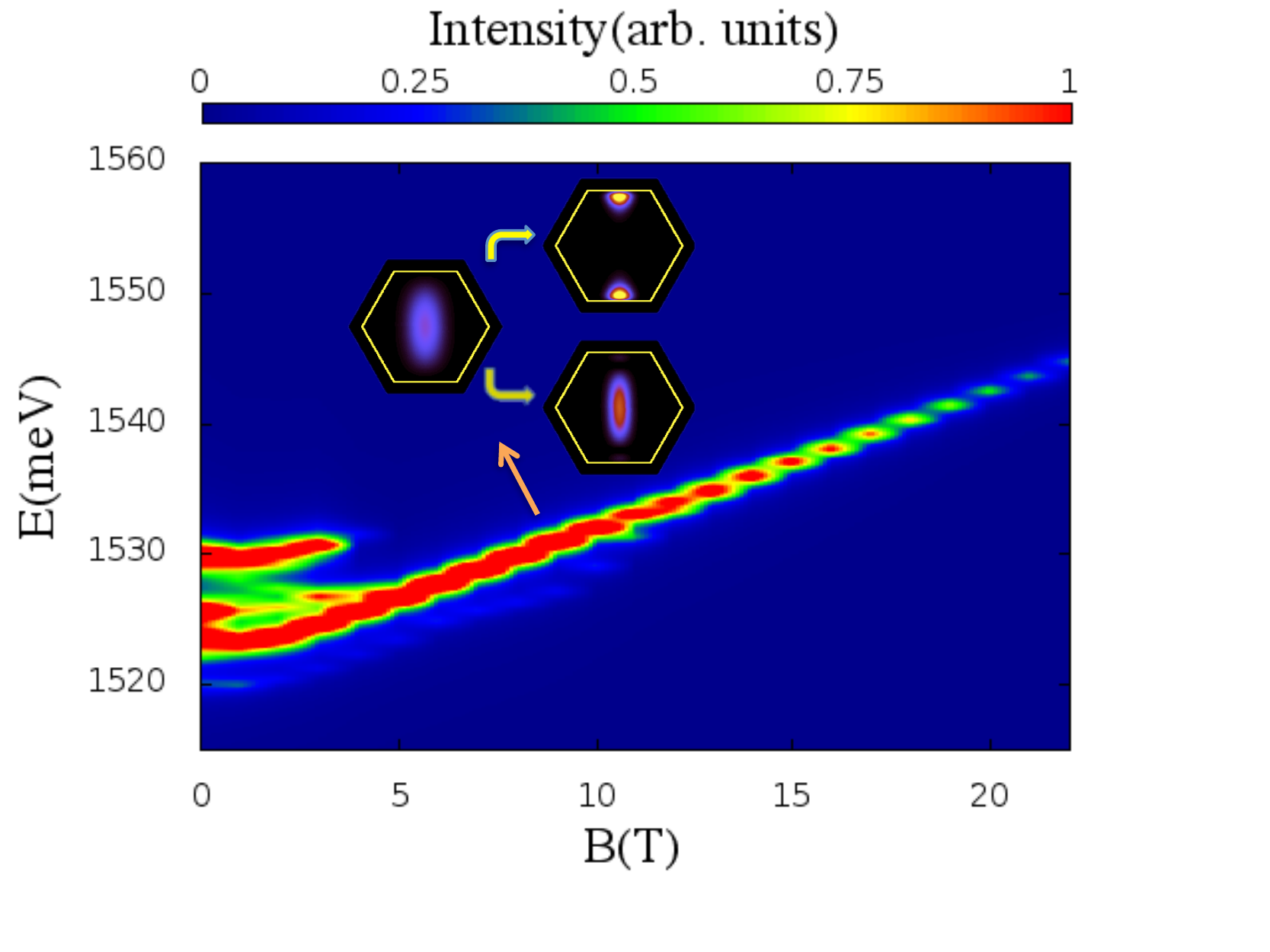}
    \caption{ \label{subbaPh8} Same as Fig.~\ref{subba}, but with $n^{\textrm{ph}}_{\textrm{l}}=10^{6}$~cm$^{-1}$. The inset in the right panel displays the squared envelope functions for the  $n=1$ electron (left) and  $n=1$ (top right)  and $n=3$ hole (bottom right) states whose recombinations yield the two peaks at  $B=10$~T.}
  \end{center}
\end{figure}
\end{widetext}

\begin{widetext}
\phantom{aa}
\begin{figure}[htpb]
  \begin{center} 
    \includegraphics[width=0.45\linewidth]{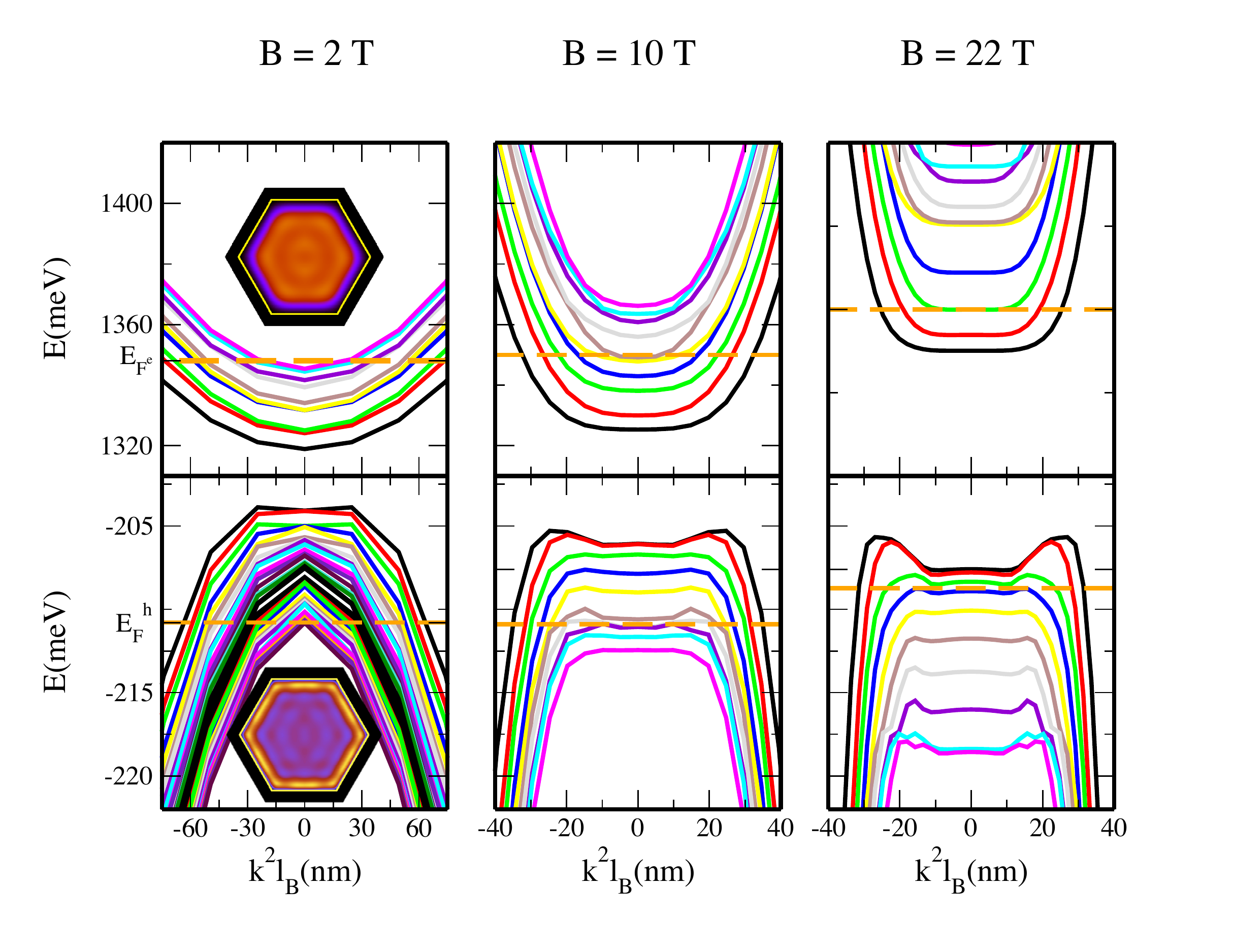}
     \includegraphics*[width=0.45\linewidth]{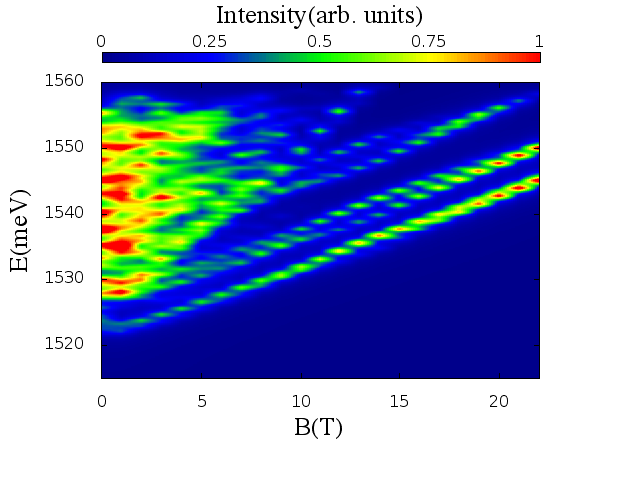}
        \caption{ \label{subbaPh9} Same as Fig.~\ref{subba}, but with $n_{\textrm{l}}^{\textrm{ph}}=10^{7}$~cm$^{-1}$. }
  \end{center}
\end{figure}
\end{widetext}

\begin{figure}[htpb]
  \begin{center} 
   \includegraphics*[width=\linewidth]{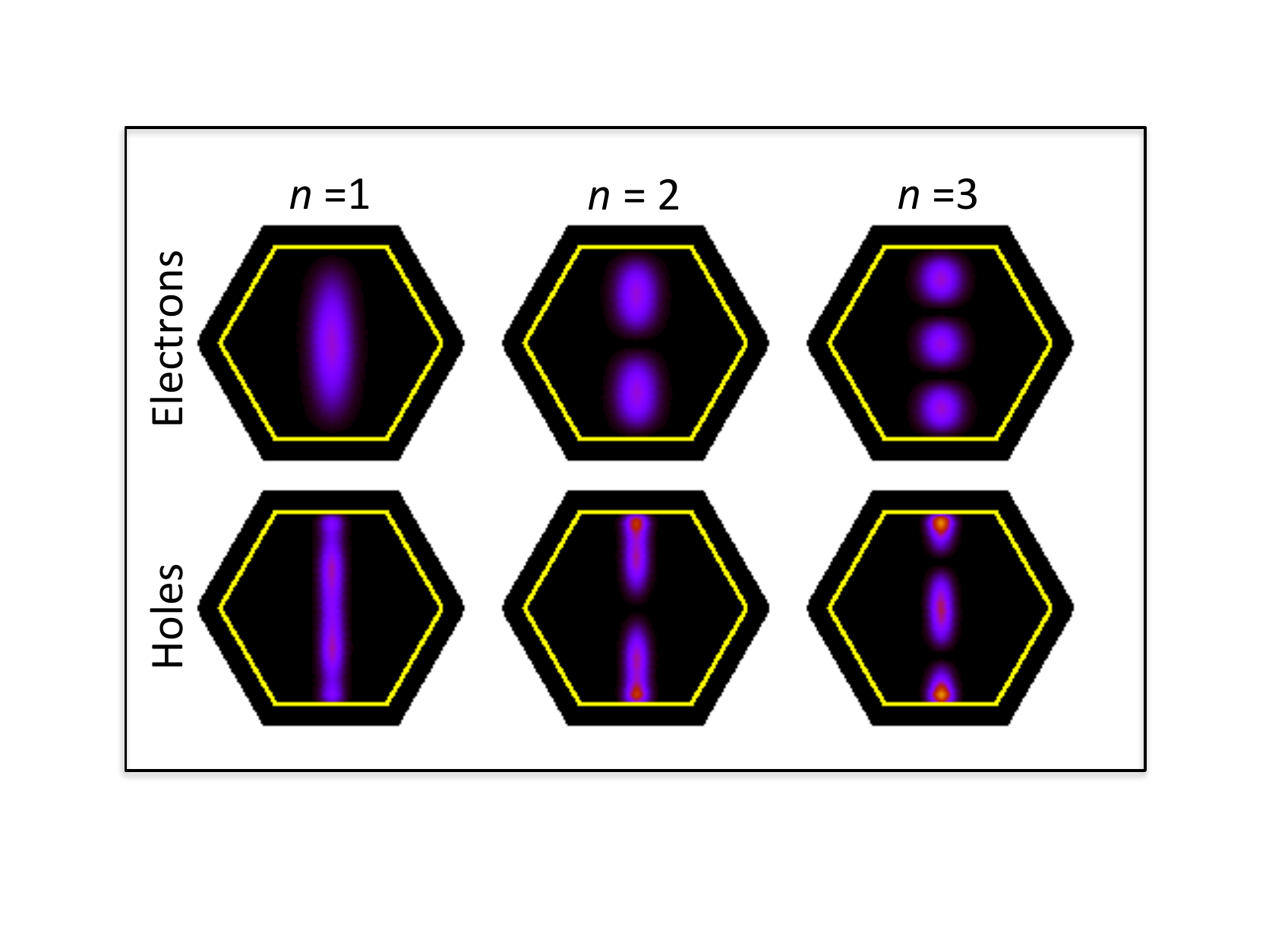}
            \caption{ \label{1e9states} 
            Square modulus of the lowest  $k=0$ electron (top) and hole (bottom) envelope function 
            at $B=22$~T for the sample of Fig.~\ref{subbaPh9}. }
  \end{center}
\end{figure}

In the large excitation power regime, namely $n^{\textrm{ph}}_{\textrm{l}}=10^{7}$~cm$^{-1}$, the density of photoexcited holes is about equal to free hole density ($0.85\times10^{7}$~cm$^{-1}$ at $B=0$, $0.84\times10^{7}$~cm$^{-1}$ at $B=22$~T).
Therefore, the electron density is now about one half the total density of holes, and a large number of both conduction and valence subbands are occupied, as shown in Fig.~\ref{subbaPh9}.
Consequently, the low field PL spectrum shown in Fig.~\ref{subbaPh9} is considerably more complex than in the previous cases, covering the range 1523~meV to 1555~meV. 
As $B$ rises, the number of peaks decreases and in the high-field region three PL emission lines can be identified, each originating from \emph{diagonal} transitions involving electron and hole states with $n=1,2,3$ (the square modulus of these wavefunctions is reported in Fig.~\ref{1e9states}).
Unlike the low-excitation power regime, the ground hole state  also spreads over the central region due to the Coulomb attraction from free electrons.
The peak at 1545 meV for $B$= 10~T related to the transition $\psi_{3,0}^e \rightarrow \psi_{3,0}^h$  decreases its intensity with the field while the other two emission lines due to  the $n=1,2$ diagonal transitions increase  unlike what observed before in the  PL spectra for lower excitation power. 
Here, the occupation number of $n=1,2$ electron and hole levels at $k=0$ is always equal to 1 for any value of $B$, and  the intensity of the emission lines is only modulated by the overlap integral which increases at larger fields.  The lowest electron and hole states exhibit the same localization: the ground states spread over the core while the
$n=2$  ones are concentrated in two lobes localized  on opposite sides. Such features are more marked for larger fields and, thus the integral overlap between the electron and holes states rises.

 \begin{widetext}
 \phantom{aa}
  \begin{figure}[htpb]
  \begin{center} 
    \includegraphics*[width=0.45\linewidth]{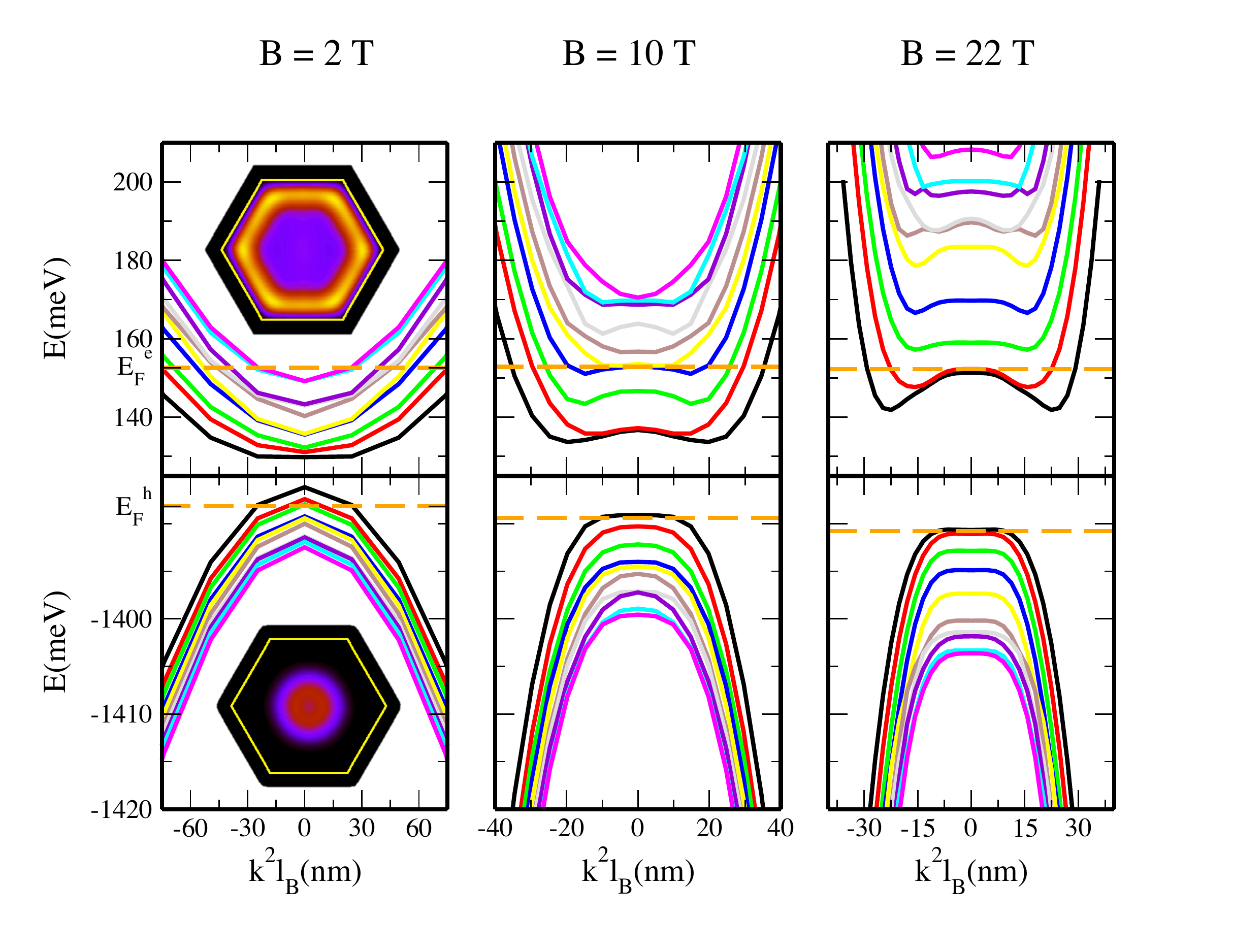}
    \includegraphics*[width=0.45\linewidth]{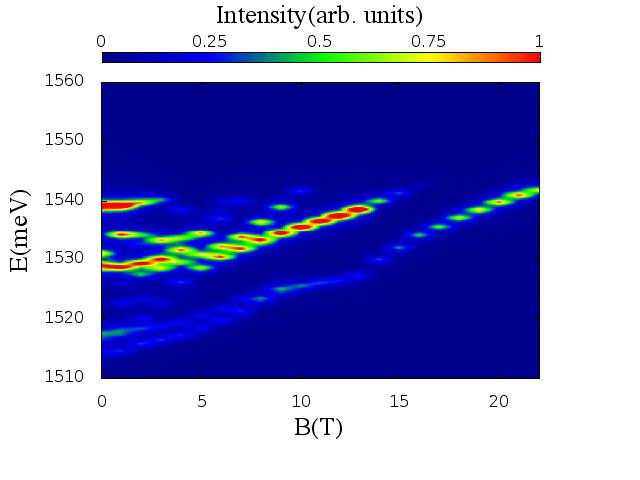} 
    \caption{ \label{subbaele} Same as in Fig.~\ref{subba} but for a $n$-doped sample with with $n_D = 4.5\times10^{19}$ and $n_{\textrm{l}}^{\textrm{ph}}=10^{6}$~cm$^{-1}$}
  \end{center}
\end{figure}
\end{widetext}

 \subsubsection{$n$-doping} \label{resndop}

We next study the fan diagrams obtained in a $n$-doped sample with $n_D=4.5\times10^{19}$~cm$^{-3}$ and a $n^{\textrm{ph}}_{\textrm{l}}=10^{6}$~cm$^{-1}$ with the same field configuration as in the previous section. 
Now the hole density is only due to photo-excitation, while the electron density due to photo-excitation is small, but not negligible, with respect to the free charge induced by $n$-doping ($9.4\times10^{6}$~cm$^{-1}$ at $B=0$, $9.5\times10^{6}$~cm$^{-1}$ at $B=22$~T). 
From this point of view, this regime corresponds to the second case analyzed in $p$-doped systems (and reported in Fig.~\ref{subbaPh8}). 
Due to the large donor density, which induces an inversion layer at the hetero-interface, in the $n$-doped case the electron subbands deviate significantly from parabolicity, while holes tend to form Landau-like and edge states.
For sufficiently large fields, the electron subbands develop minima which correspond to states localized at the flanks. 

\begin{figure}[htpb]
  \begin{center} 
  \includegraphics*[width=\linewidth]{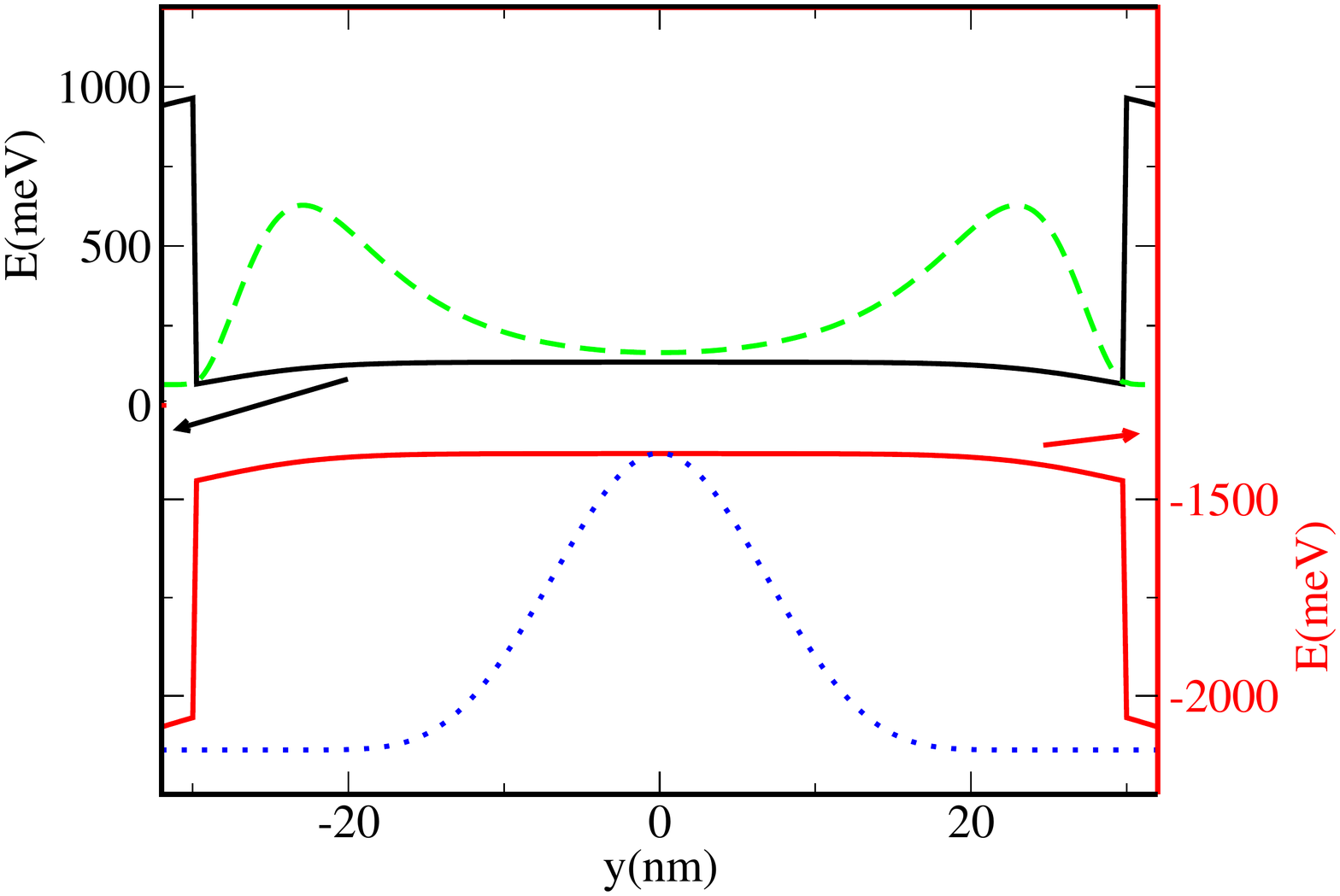}
  \includegraphics[width=0.9\linewidth]{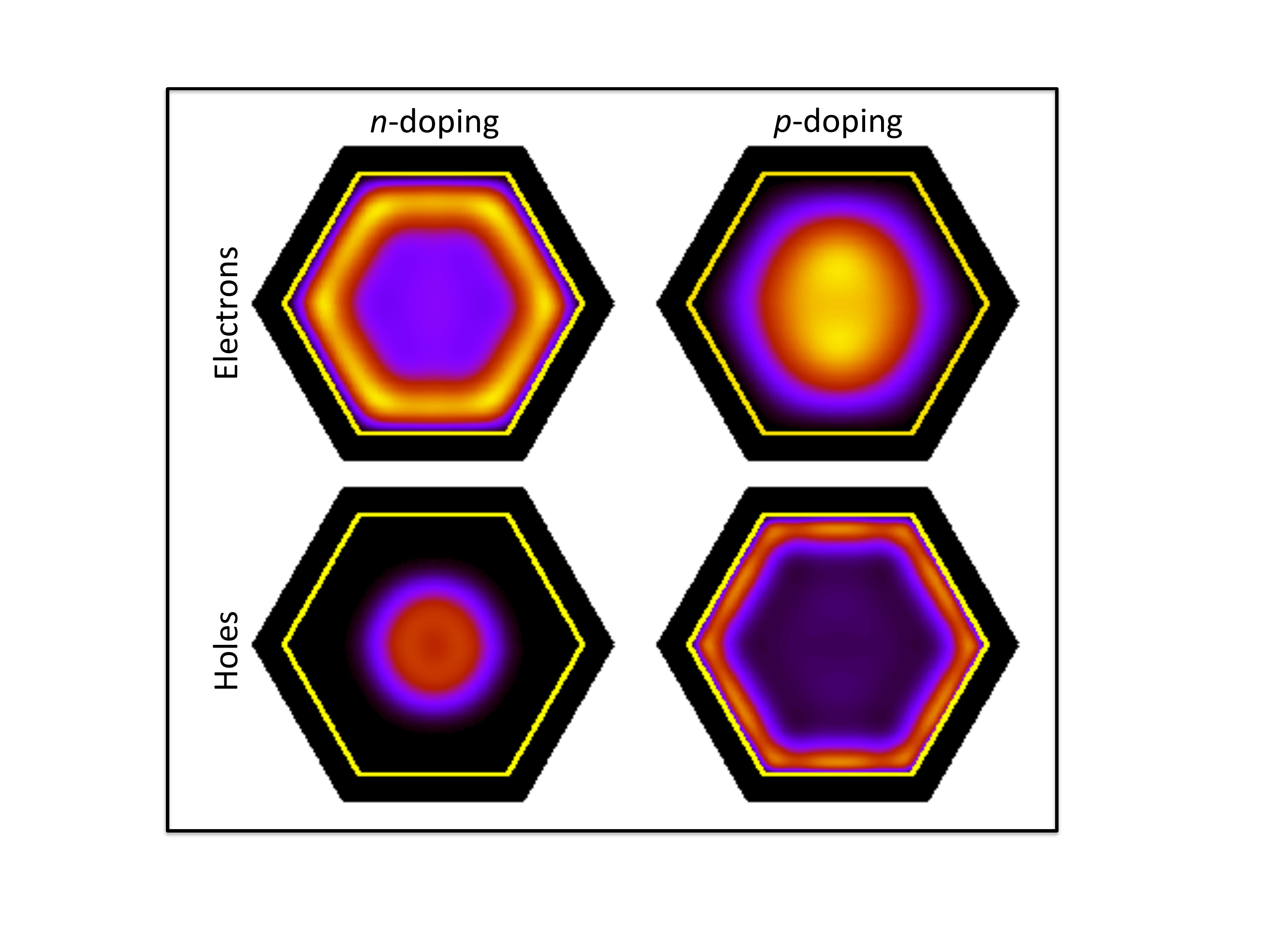}
    \caption{ \label{chaLoc}  Top: self consistent conduction (black line)  and valence (red line) band profile  and the square modulus of  the 
    $n=1$ electron (green dashed line) and hole (blue dotted line) wavefunction along the $y$-axis at $B=0$. Bottom: electron and hole charge densities with $n_{\textrm{l}}^{\textrm{Ph}}=10^{6}$~cm$^{-1}$ at $B=2$~T for  $n_D = 4.5\times10^{19}$~cm$^{-3}$ (left panels) and $n_A=2.5\times10^{19}$~cm$^{-3}$ (right panels).}
  \end{center}
\end{figure}

The right panel of Fig.~\ref{subbaele} displays the corresponding PL spectra. For $B=0$ we observe several emission lines. 
Note that the lowest one is at 1515~meV, below the bulk GaAs band gap. Indeed, as shown in Fig.~\ref{chaLoc} (top panel),  the hole ground state spreads over the GaAs core while the electron ground state is localized at the inner GaAs/AlAs interfaces depleting  the central zone. 
The bending  of the bands due to electrostatic potential mostly generated by the dopants is more pronounced  in the regions farther  away from the  central core. 
Therefore, the down shift of electron levels is generally larger than for hole states, and, as a consequence, the  energy of ground e-h state lies below the bulk GaAs band gap. 
Such a behavior was not observed in the $p$-doping regime since in that case the band bending was less pronounced due to the smaller density of dopants ($n_A=2.5\times10^{19}$~cm$^{-3}$ \textit{vs} $n_D=4.5\times10^{19}$~cm$^{-3}$).
 
In this sample, the number of transitions at low fields is larger than a for $p$-doped sample in the corresponding regime of photo-excitation power, although the number of subbands occupied by the minority carriers is similar. 
This is due to the different localization of the majority charge which, for $n$-type doping (although mainly localized at the hetero-interface, as seen in the bottom panel of Fig.~\ref{chaLoc}), extends to the core and overlaps with the hole ground state. 
As the magnetic field increases, the number of emission lines decreases due to hole state depopulation. At $B=10$~T, only the hole ground state and few electron subbands are occupied while the Fermi level is pinned, as shown in left panels of Fig.~\ref{subbaele}. In the intermediate fields regime we identify three main peaks originating from the recombination of the hole ground state with different electron levels. 

\begin{figure}[htpb]
  \begin{center} 
  \includegraphics*[width=0.45\linewidth]{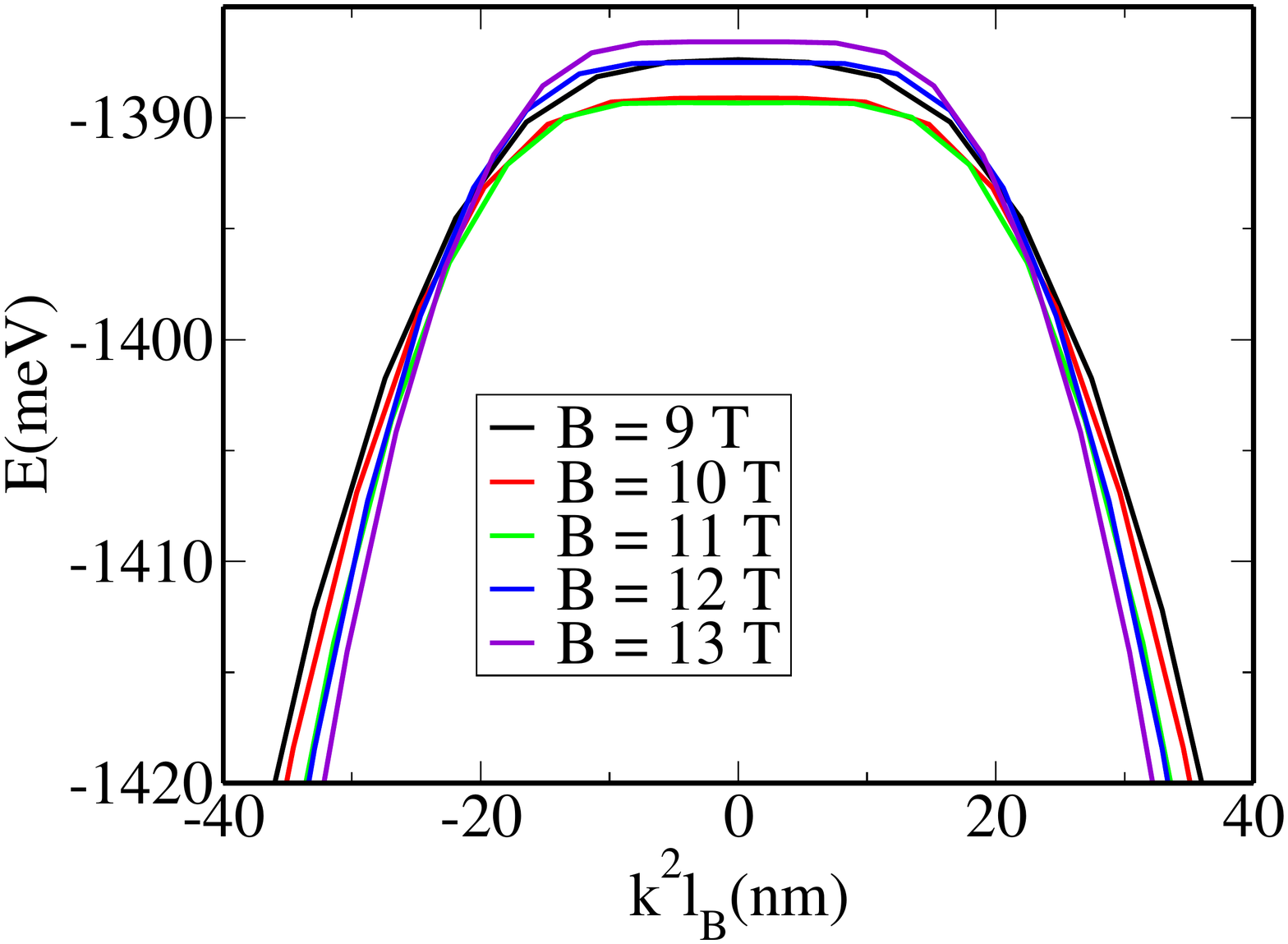}
   \includegraphics*[width=0.45\linewidth]{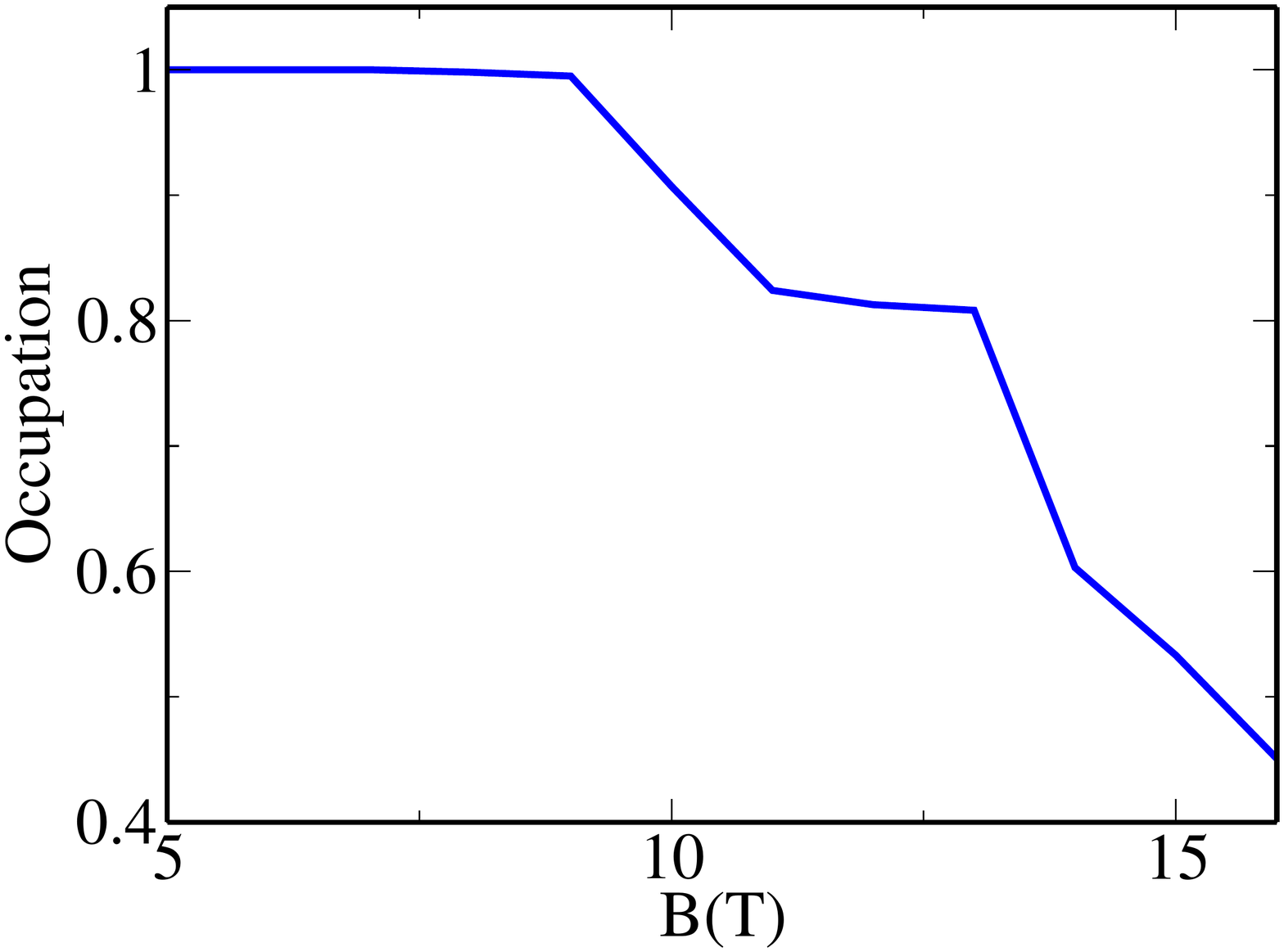}
    \includegraphics*[width=\linewidth]{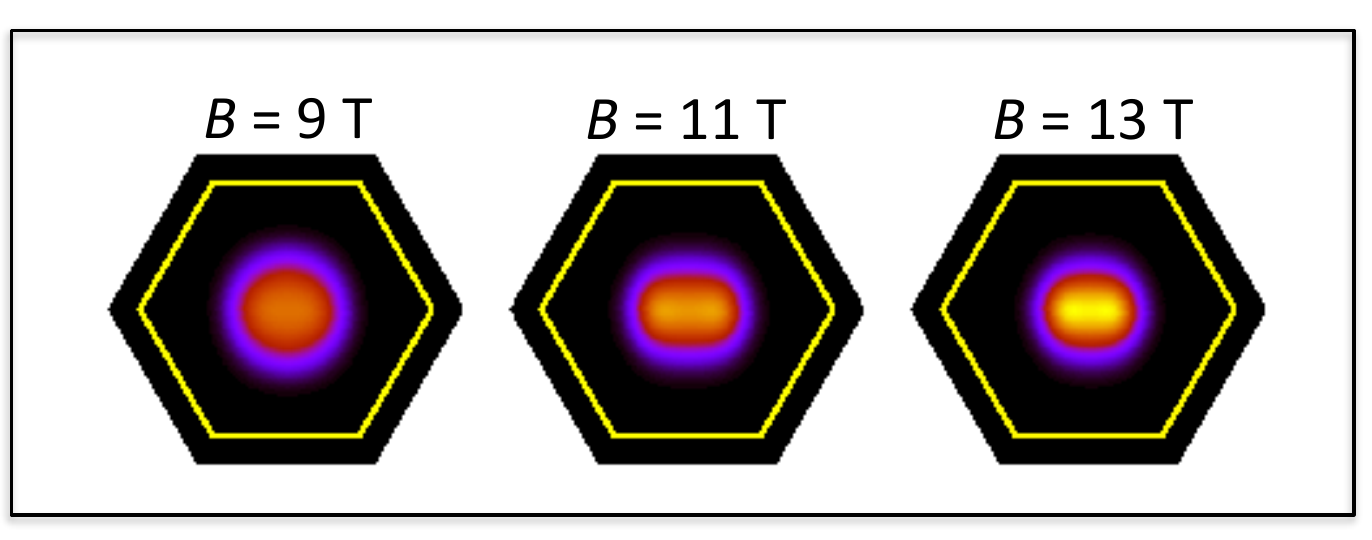}
    \caption{ \label{ELEC}  Top left: $n=1$ hole subband \textit{vs} $k l^2_B$ evaluated for $B$ ranging from 9 to 13~T. Top right: occupation of $n=1$ hole state at $k=0$ \textit{vs} $B$. Bottom: hole gas density at selected fields, as indicated. }
  \end{center}
\end{figure}

The emission energy due to the e-h ground state exhibits a flat region between $B=9$ and $12$~T. For $B > 9$~T the occupation of the $n=1$ hole state at $k=0$ decreases significantly since the lowest subband flattens and states with $k\neq 0$ are increasingly populated at the expense of the $k=0$ state, as shown in the top panels of Fig.~\ref{ELEC}. 
Since finite $k$ states are localized away from the NW axis, the field induces a restructuring of the hole density   which becomes anisotropic (see the bottom panel of Fig.~\ref{ELEC}). 

For the high-field regime, only one peak survives, due to the recombination of the ground e-h state.

\subsection{Anisotropy} \label{subani}

Carriers are either homogeneously distributed around the hetero-interface or form 1D and 2D channels according to the device parameters.
In the latter case, we expect the direction of the field to have a remarkable impact.
Therefore, we next compare PL spectra with the field oriented in two non-equivalent directions, namely perpendicular to the top/bottom facets ($\theta = \pi/2$) and along a maximal diameter ($\theta = \pi/3$). 
We assume $p$-doping with $n_A=2.5 \times 10^{19}$~cm$^{-3}$ at the intermediate excitation power regime, that is $n_{\textrm{l}}^{\textrm{Ph}}=10^{6}$~cm$^{-1}$. Left panels of Fig.~\ref{anisitropy} display the PL spectra evaluated  for the two configurations at different values of the field intensity.
\begin{widetext}
\phantom{aa}
\begin{figure}[htpb]
  \begin{center} 
    \includegraphics*[width=\linewidth]{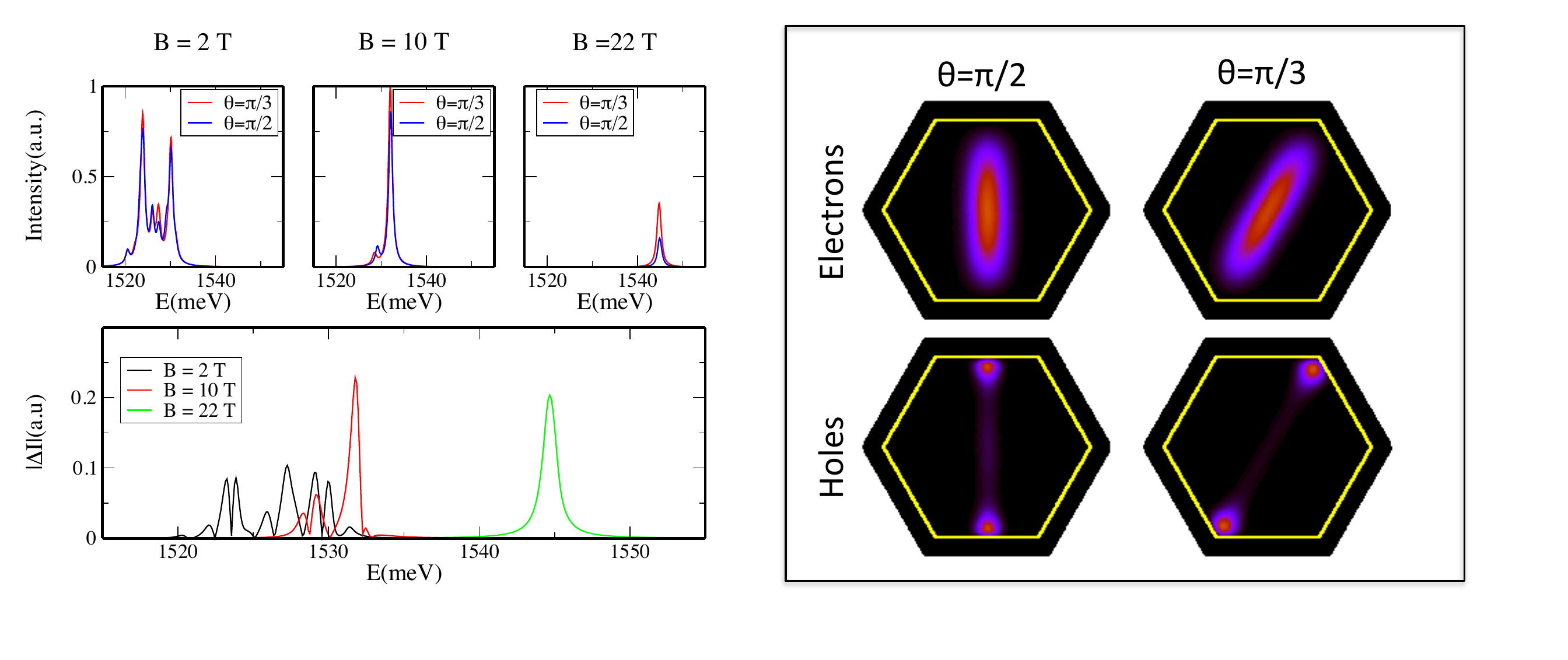}
           \caption{ \label{anisitropy} Left: PL spectra in different directions (top) and absolute difference between the PL spectra (bottom) at selected fields, as indicated. Right: square modulus of the $n=1$ hole and electron envelope functions at $k=0$ and $B=22$~T for the two field orientations.}
  \end{center}
\end{figure}
\end{widetext}
The spectra exhibit the same qualitative behavior, with several emission lines at low $B$ which reduce to one peak (ground e-h state) at larger fields. 
However, the emission intensity is anisotropic, especially at high field intensity. 
At $B = 22$~T, the peak intensity at $\theta = \pi/3$ is about twice as strong as the one at $\theta = \pi/2$. 
This is due to the different localization of the $n=1$ electron and hole states at $k=0$, shown in the right panel of Fig.~\ref{anisitropy}. 
The hole state shows two lobes localized by the strong carrier-carrier repulsion towards the bottom and top facets (at $\theta = \pi/2$) or corners (at $\theta = \pi/3$).
Since the facet-to-facet distance is the shorter, due to Coulomb interaction the  energy of the state in the first configuration turns out to be a few tenths of meV greater than the one with the field along the maximal diameter.
As a consequence, its occupation is smaller.
The electron states elongate in the field direction and, also in this case, the energy level obtained at $\theta = \pi/2$ turns out to be greater than the one found at $\theta = \pi/3$, and this yields a discrepancy in the occupation of the states. 
Thus, the higher population of both $n=1$ hole and electron states leads to the stronger intensity of the ground e-h state emission line at  $\theta = \pi/3$ with respect to the one found at $\theta = \pi/2$.

\section{Summary and conclusions}\label{conclu}


We have investigated theoretically the magneto-optical properties of hole and electron gases in GaAs/AlAs core-multishell NWs which are either $n$- or $p$-doped and subject to a transverse magnetic field. The tailoring of PL spectra by the field has been rationalized in different doping regimes.
Calculations were performed by means of a self-consistent mean-field approach, with the inclusion of photogenerated hole-electron pairs that add to the free carriers induced by the modulation doping.

Our results indicate that  the behavior of  magneto-PL spectra offers a valuable guideline to estimate the density of the doping in CMS NWs. 
In fact, when the density of free carriers due to the doping is greater than the one of  photogenerated charges, the spectra are quite simple and only the emission line due to the ground e-h state survives at high fields. Differently, when the excitation power is larger and the density of photoexcited carriers is comparable or higher with respect to the doping-induced free charge, several PL peaks are found, that can be ascribed to recombinations from specific excited states, whose character is, in turn, strongly modified by the magnetic field.

We found a remarkable difference between samples with either electron or holes as majority carriers. 
Indeed, at low magnetic fields and low photoexcitation power, the qualitative behavior of PL spectra for $n$- and $p$-doped sample is quite similar although, in the case of $n$-doping, the number of emissions lines is greater than the one found for the $p$-doped sample. This is a consequence of the different localization pattern of the majority carriers.
Still in the low-field regime, when the density of photo-created hole-electron pairs is significant, the emission lines exhibit a diamagnetic shift, in both samples.
This shift becomes linear for higher magnetic fields. However, the main peak, due to the ground e-h state, deviates from the expected behavior at low values of density of photogenerated hole-electron pairs. Specifically, in the intermediate field regime, it exhibits a kink ($p$-doping) or a flat region ($n$-doping). 
In both cases this is due to the rearrangement of the gas of holes, whose heavier effective mass eases their localization near the interface with respect to the electrons. 
In the high power excitation regime, the large Coulomb interaction prevents the hole gas rearrangement.

The differences in the magneto-PL spectra illustrated in this work are a valuable tool to discriminate not only the free-carrier density regime and type, but also the relative density of doping- and photoexcitation-induced carriers. This, together with the assessment of the free-carrier spatial distribution, is pivotal in the characterization and engineering of CMS NWs.

\begin{acknowledgments} 
Numerical simulations were performed at CINECA within the Iscra C project MPL-CSNW. We acknowledge partial financial  support from University of Modena and Reggio Emilia, with Grant ``Nano- and emerging materials and systems for sustainable technologies." M.~R. acknowledges financial support from APOSTD/2013/052
Generalitat Valenciana Vali+d Grant,.
\end{acknowledgments}


%

\end{document}